# Machine Learning in Magnetic Resonance Imaging: Image Reconstruction

Javier Montalt-Tordera (javier.tordera.17@ucl.ac.uk) [a]
Vivek Muthurangu (vivek.muthurangu@ucl.ac.uk) [a]
Andreas Hauptmann (a.hauptmann@ucl.ac.uk) [b,c]
Jennifer Anne Steeden (jennifer.steeden@ucl.ac.uk) [a]

a. UCL Centre for Cardiovascular Imaging, University College London, London. WC1N 1EH. United Kingdom
b. University of Oulu, Research Unit of Mathematical Sciences, Oulu. Finland
c. Department of Computer Science, University College London, London. WC1E 6BT. United Kingdom

**Corresponding Author:** Jennifer Steeden
UCL Centre for Cardiovascular Imaging, Institute of Cardiovascular Science, 30 Guildford Street, London. WC1N 1EH
jennifer.steeden@ucl.ac.uk
Tel: +44 (0)207 762 6834
Fax: +44 (0)207 813 8263

**Declarations of interest:** None

## Abstract

Magnetic Resonance Imaging (MRI) plays a vital role in diagnosis, management and monitoring of many diseases. However, it is an inherently slow imaging technique. Over the last 20 years, parallel imaging, temporal encoding and compressed sensing have enabled substantial speed-ups in the acquisition of MRI data, by accurately recovering missing lines of *k*-space data. However, clinical uptake of vastly accelerated acquisitions has been limited, in particular in compressed sensing, due to the time-consuming nature of the reconstructions and unnatural looking images. Following the success of machine learning in a wide range of imaging tasks, there has been a recent explosion in the use of machine learning in the field of MRI image reconstruction. A wide range of approaches have been proposed, which can be applied in *k*-space and/or image-space. Promising results have been demonstrated from a range of methods, enabling natural looking images and rapid computation. In this review article we summarize the current machine learning approaches used in MRI reconstruction, discuss their drawbacks, clinical applications, and current trends.

# 1  Introduction

## 1.1  The Image Reconstruction Problem

Magnetic Resonance Imaging (MRI) is extensively employed in medical diagnosis and is a reference standard in many applications. However, it has a significant drawback: the inherently slow nature of data acquisition. The MRI signal is generated by the nuclei of hydrogen atoms as they interact with external electromagnetic fields. However, an MRI scanner cannot measure spatially dependent signals (i.e. images) directly. Rather, the spatial dependence is encoded into the frequency and phase of the MRI signal. This encoding process is inherently sequential, which leads to long acquisition times. Ultimately, a spatial frequency map is obtained, which is referred to as *k*-space. In the simple case, the inverse Fourier transform (iFT) can then be used to reconstruct the *k*-space data into clinically interpretable images.

Due to the sequential nature of MRI scanning, acquisition time is roughly proportional to the number of *k*-space samples collected. Therefore, it is desirable to collect as few samples as possible. However, if the sampling rate is reduced below that required by the Nyquist criterion, aliasing artefacts will appear in the image.

In general terms, the image reconstruction can be formulated as the following inverse problem:

$$y = Ax + \epsilon \qquad (1)$$

where $y$ is the measured *k*-space data, $A$ is the system matrix, $x$ is the image and $\epsilon$ is a random noise term. When *k*-space data is undersampled and noise corrupted, the inverse problem in Equation 1 is ill-posed: a solution might not exist, infinite solutions might exist, and it may be unstable with respect to measurement errors. As a result, direct inversion of $A$ is generally not possible. Instead, an optimal solution in the least-squares sense may be obtained by recasting the problem as the following minimization:

$$\hat{x} = \underset{x}{\mathrm{argmin}}\, \frac{1}{2} \|Ax - y\|_2^2 \qquad (2)$$

Much research effort has been devoted to image reconstruction from an undersampled *k*-space over the last few decades. Two broad technologies stand out for their importance and deserve a brief overview here, namely parallel imaging and compressed sensing. These enable substantial reductions in acquisition time while preserving image quality.

Parallel imaging techniques exploit multi-channel receiver arrays to compensate for the undersampling of *k*-space. This is enabled by the fact that receiver coils exhibit spatially varying responses, which can be leveraged to unfold aliased images or estimate missing *k*-space samples. Parallel imaging techniques, such as SENSE (Sensitivity Encoding) [1] and GRAPPA (Generalized Autocalibrating Partial Parallel Acquisition) [2], enjoy tremendous success and are routinely used in the clinical environment. However, increasing acceleration factors lead to signal-to-noise ratio (SNR) losses, which in practice limits the achievable acceleration.

Compressed sensing (CS) [3] enables the reconstruction of subsampled signals provided that the signal to be reconstructed is sparse in some domain. A signal is sparse if it contains few non-zero elements compared to its size. MRI images are not typically sparse. However, like most natural images, they contain many redundancies, and have sparse representations in other domains such as the finite difference or wavelet domain. The expectation that the solution be sparse in some domain can be incorporated into the optimization problem in Equation 2 as a regularization term:

$$\hat{x} = \underset{x}{\mathrm{argmin}}\, \frac{1}{2}\|Ax - y\|_2^2 + \lambda \|Dx\|_1 \qquad (3)$$

where $D$ is the sparsifying transform, mapping a redundant image to its sparse representation, $\|.\|_1$ is the $\ell_1$-norm and $\lambda$ is a regularization parameter. The first term in this equation serves to enforce that the solution is consistent with the measured data, while the second term favors solutions that are sparse in the transform domain. The parameter $\lambda$ balances both terms and can be tuned to optimize image quality.

Compressed sensing is not only based on the expectation that the correct solution of Equation 3 is sparse in the transform domain, but also that the aliased solutions are not sparse. This translates into another requirement: that the aliasing artefact is incoherent, i.e. that it resembles noise. This can be satisfied in MRI by using non-regular or pseudo-random sampling patterns, within the hardware constraints.

Compressed sensing, which can be readily combined with parallel imaging, has enabled high acceleration factors. However, clinical translation has been complicated by several factors. First, the non-linear iterative reconstruction often takes too long or requires computational resources not currently available in most clinical services. Second, images have been reported as looking unnatural and blocky. Finally, tuning of the regularization term is an empirical process that often depends on the specific application and may be different for each patient.

In the general case, aliasing and signal degradation cannot be avoided when sampling below the Nyquist rate, as per the Nyquist-Shannon theorem. Reconstruction methods for accelerated MRI rely on some form of prior information or additional constraints on the reconstructed signal. However, the priors used in parallel imaging and compressed sensing are often crude, and more representative priors have the potential to improve current reconstruction techniques. However, designing such priors by hand is difficult. Artificial intelligence, on the other hand, excels at discovering patterns in data. Therefore, it is the ideal tool to inject the knowledge provided by historic MRI data into the image reconstruction.

The following section is a very brief overview of deep learning, the subset of artificial intelligence most relevant to MRI.

## 1.2 Deep Learning

Artificial neural networks are a class of machine learning algorithm that apply a series of cascaded layers (where each layer consists of a series of connected nodes, or neurons), mapping inputs to outputs. Each of these layers receives inputs, performs an operation and returns an output, which is then passed to the next layer. These layers have two crucial properties. First, the majority of these layers perform non-linear operations (often a linear transformation followed by a non-linearity), which when combined can represent very complex functions. Second, the layers have trainable parameters, i.e. parameters which are not fixed or designed, but optimized during a training process. During the training process, the parameters are iteratively adjusted by an optimization algorithm in order to minimize a loss function for a given set of training data. As it *learns*, the network approximates the mapping from inputs to outputs.

Deep neural networks are artificial neural networks which have multiple layers, where in general, the deeper a neural network is, the higher its representational power. The use of deep neural networks in order to discover mappings or representations is referred to as deep learning. Some special types of neural networks deserve mention for their importance in MRI. Convolutional neural networks (CNN's) are those primarily based on shift-invariant convolutional layers, where the trainable parameters are a set of convolutional kernels which are translated along the image dimensions in a sliding window fashion. They have several properties which make them ideally suited to image processing, including the ability to encode local relationships, and that they are agnostic to image size [4]. Furthermore, a commonly used principle is based on recurrent neural networks (RNN), which are designed to process sequences of inputs. These maintain a *hidden state*, which acts like a memory about previous inputs in the sequence.

It is essential that the network contains non-linearities, also called activations. The most commonly used non-linearity is the rectified linear unit, or ReLU, which is a piecewise linear function that returns the input value for positive inputs and zero for negative inputs. This function has become the default choice in many deep learning models because it often outperforms more complex activations and leads to models which are easier to train, due to its beneficial gradient properties. Nevertheless, other activations such as sigmoid or extensions such as leaky ReLU (including negative values) are sometimes used.

In general, machine learning problems can be formulated in a supervised or unsupervised manner. Supervised learning uses known ground-truth data to learn a mapping between data pairs, whereas unsupervised learning infers structures within the sample without labelled outputs. Currently, most applications in MRI reconstruction use supervised learning, however unsupervised techniques remain an area of active research. There have been many approaches to machine learning MRI reconstruction (both in terms of supervised and unsupervised techniques), including methods which work in image-space, those which work in *k*-space, those which operate in different domains, those that learn the direct mapping from *k*-space to image-space, and unrolled optimization methods (Figure 1).

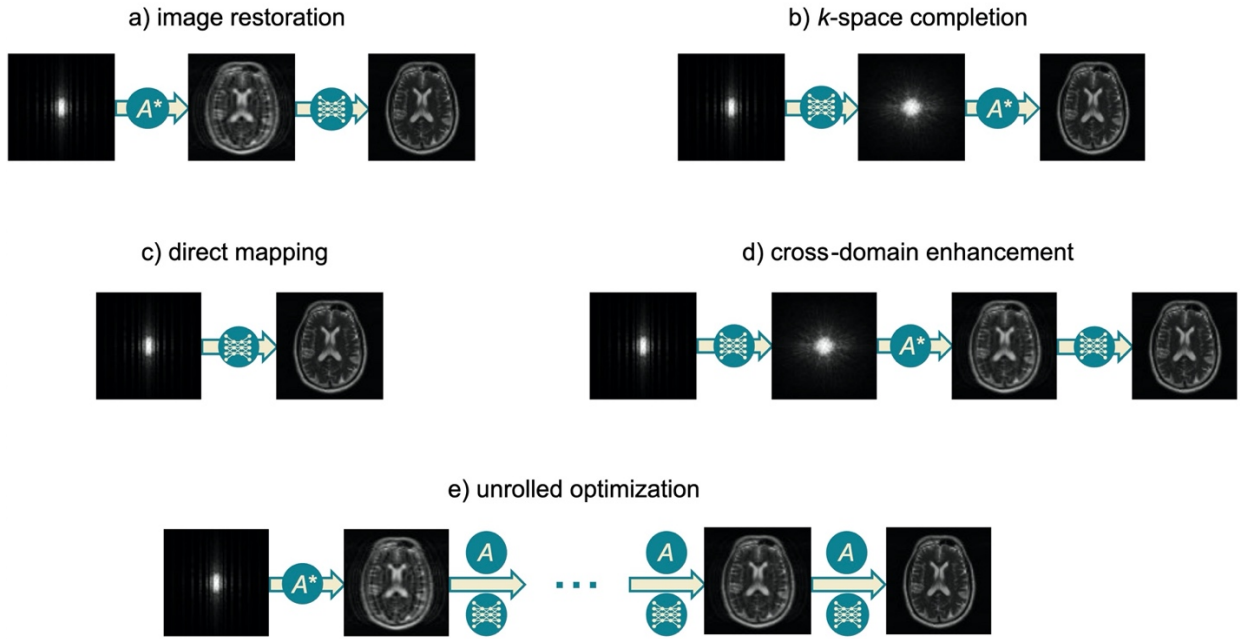

*Figure 1: Image reconstruction methods can be roughly classified into five categories: a) image restoration, b) k-space completion, c) direct mapping, d) cross-domain enhancement, and e) unrolled optimization, depending on how neural networks are used. $A$ is the image formation model, and $A^*$ is the adjoint operator.*

## 2   Supervised Machine Learning

In supervised learning the ML algorithm learns a function that maps the input to an output from a training data set, consisting of paired input and output images. This requires a *gold-standard* fully sampled data set (the desired output), with paired undersampled data (the input). These approaches require a qualitative metric, or *loss function*, which is used to evaluate how close the current output of the network is to the target image. The most commonly used loss functions are pixel-wise Mean Squared Error (MSE, $\ell_2$-loss) and Mean Absolute Error (MAE, $\ell_1$-loss). However, these metrics do not reflect a radiologists' perspective well [5], and are generally not good at representing small structures. Development of new loss functions, including feature losses, remains an area of active research [6-8].

## 2.1 Image Restoration Methods

Image restoration techniques are those that operate in the image domain only (see Figure 1-a). These methods relate closely to general image problems in non-medical contexts, including image processing. As a result, they can directly benefit from and contribute to the rich body of literature on CNN-based image enhancement, including de-noising and super-resolution. The first applications of machine learning to MRI reconstruction were based on image restoration methods [9]. A popular network in these methods is the convolutional encoder-decoder architecture with skip connections, also known as *U-Net* [10]. It consists of an encoder path, with multiple down-sampling steps with increasing number of channels, followed by a decoder path, with multiple up-sampling layers with decreasing number of channels (see Figure 2). In addition, skip connections are added between encoding and decoding steps operating at the same scale, so that each decoding step receives the concatenation of the previous decoding step and the corresponding encoding step as its inputs.

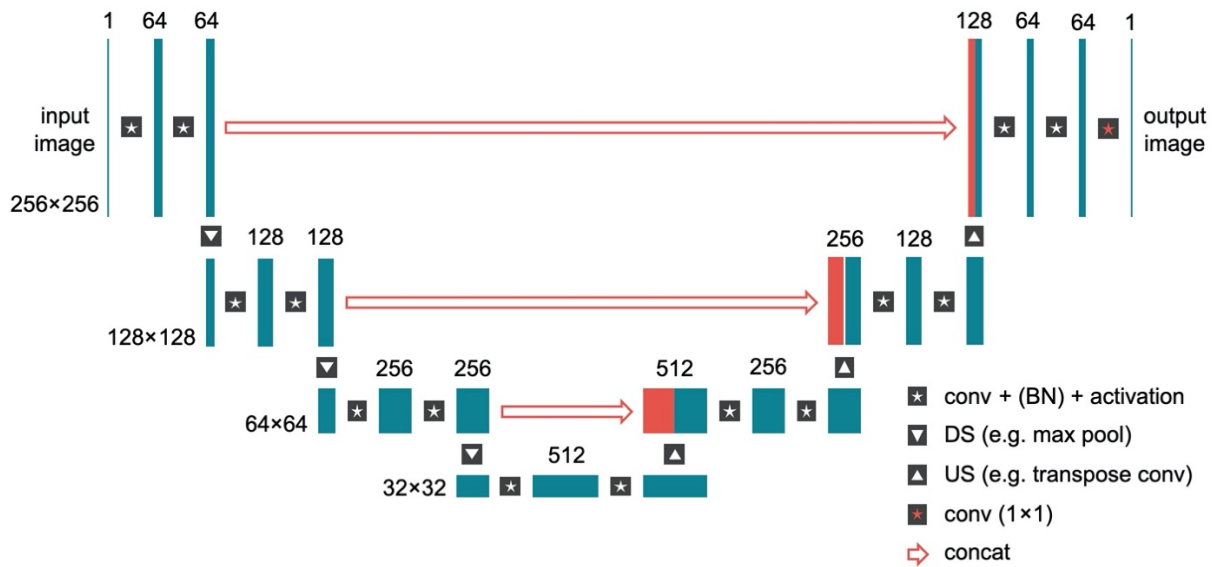

*Figure 2: A possible variant of the commonly used U-Net architecture (encoder-decoder with skip connections). The height of the blocks represents changes in spatial resolution while the width represents the number of channels. Example values are shown for the sake of clarity. Downsampling (DS) is often done using max-pooling layers. Upsampling (US) can be achieved using up-sampling or transpose convolution layers. The activation is often a ReLU. Batch normalization (BN) layers are sometimes added to stabilize training. A special 1×1 convolution, also called bottleneck layer, is often used at the end to reduce the channel dimension. Concatenation operations relay the features at each scale of the encoder path to the corresponding scale in the decoder path. The number of scales may vary.*

Undersampling of *k*-space results in aliasing artefacts in the reconstructed images, which are dependent on the trajectory and undersampling pattern. Where the undersampling is performed in a non-uniform manner, the resultant artefacts are incoherent and noise-like. Therefore, it is possible to train a machine learning network to remove such artefacts in a similar manner to image de-noising. It has been shown that it is possible to perform de-aliasing from data acquired using a random undersampling scheme in the phase direction of 2D images [11]. In these applications it has been shown that there is a benefit to training a CNN to learn the residual (i.e. the aliasing artefact) rather than the corrected image, because the residual has lower topological complexity [11]. This has been expanded to 2.5D (where time is included in the channel dimension) using a 2D Poisson disk sampling mask [12], as well as to golden-angle radial sampling using a 2D CNN with spatio-temporal slices [13], and using a 3D U-Net (2D plus time) [14]. It has also been used on complex data, to de-alias phase contrast MRI images [15].

Another group of de-aliasing methods uses generative adversarial networks (GAN's) [16]. GAN's consist of two subnetworks: a generator, which produces images based on some input; and a discriminator, which attempts to distinguish the generator output from ground-truth images. During training, the generator learns to produce realistic images so as to deceive the discriminator, by minimizing an adversarial loss. In MRI reconstruction, this adversarial loss is typically combined with a pixel-wise distance loss (such as the $\ell_1$- or $\ell_2$-norm) to stabilize training and ensure consistency with the ground-truth image. Discriminator networks are typically vanilla CNN classifiers; however, there is more variability in the types of generator networks used. Some examples include; Deep De-Aliasing Generative Adversarial Networks (DAGAN) [17] uses a U-Net architecture for the generator network. RefineGAN [18] uses a cascade of two U-Nets, with the first performing the reconstruction and the second refining this result. GANCS (GAN for compressive sensing) [19] uses a deep residual network (ResNet) [20] as the generator, and also includes an affine projection operator for data consistency.

An alternative method to speed up MRI imaging is to acquire lower resolution data, using a smaller base matrix. It is then possible to apply the image enhancement method; super-resolution (SR), which attempts to predict high-frequency details from low-resolution images. Because SR can be simply applied as a post-processing step, there have been many applications of super-resolution in MRI reconstruction. Simple network structures include Super-Resolution Convolutional Neural Networks (SRCNN) [21] which learn end-to-end mapping. This has been applied to 2D brain MRI images [22], and extended to 3D brain images [23], as well as dynamic cardiac MRI data [24]. This has further been improved through the use of 3D densely-connected blocks (DCSRN) [25] and dense connections with deconvolution layers (DDSR) [26], as well as residual U-Net structures [27]. Most studies demonstrate good results with two or three-fold down sampling. A full review of the use of machine learning super-resolution in medical imaging can be found [28].

## 2.2  *k*-space Methods

Machine learning networks have been trained to perform *k*-space enhancement (see Figure 1-b), in a supervised manner, similarly to GRAPPA. Some approaches use large training databases without the need for explicit coil-sensitivity information, whereas others learn the relationship between coil elements from a small amount of fully sampled reference data (the auto-calibration signal, ACS).

DeepSPIRiT [29] uses CNN's to interpolate undersampled multi-coil *k*-space data. It is based on the SPIRiT (iterative self-consistent parallel imaging reconstruction) algorithm [30], which is a generalizable coil-by-coil reconstruction based on self-consistency with the acquisition data. To enable DeepSPIRiT to be used with different hardware configurations and different numbers/types of coils, the data is first normalized using coil compression with principal component analysis (PCA) [31]. This places the dominant virtual sensitivity map in the first channel, and the second dominant in the second channel, etc. Different regions of *k*-space are trained separately in a multi-resolution approach, using a large database without the need for explicit coil sensitivity maps or reference data. Where multiple contiguous slices are available, spatially adjacent slices can be used as multi-channel input to improve the accuracy; adaptive convolutional neural networks for *k*-space data interpolation (ACNN-*k*-Space) [32].

Alternative methods exploit the low-rank of the MRI signal, similarly to ALOHA (annihilating filter based low-rank Hankel matrix) [33]. It has been shown that it is possible to train a U-Net using a large database, which exploits the efficient signal representation in *k*-space [34, 35].

Other machine learning approaches are more closely related to the parallel imaging technique, GRAPPA. RAKI (Scan-specific robust artificial-neural-networks for k-space interpolation) [36] is trained on the ACS data to learn the non-linear relationship between coil elements. Therefore, RAKI does not require a large training database, but instead the neural networks are trained using the ACS data from the scan itself. This means that the network must be trained for each scan. The resulting RAKI networks have been shown to lead to a reduction in noise amplification compared to GRAPPA. The use of arbitrary sampling patterns are possible with the use of self-consistent RAKI (sRAKI) [37]. Other advances includes residual RAKI (rRAKI) [38], which uses a residual CNN to simultaneously approximate a linear convolutional operator and a non-linear component that compensates for noise amplification artefacts. Furthermore, RAKI has been combined with LORAKS (Low-rank modelling of local *k*-space neighborhoods) [39], in a method called LORAKI [40]. LORAKI uses an auto-calibrated scan-specific convolutional RNN, which simultaneously incorporates support, phase, and parallel imaging constraints.

## 2.3   Direct Mapping

A few studies have shown the possibility of directly learning the transform between the undersampled *k*-space data and the uncorrupted images (see Figure 1-c). These end-to-end reconstructions have the potential to mitigate against errors caused by field inhomogeneity, eddy current effects, phase distortions, and re-gridding.

AUTOMAP (automated transform by manifold approximation) [41] was trained using a large database of paired synthetic undersampled *k*-space data (input), and reconstructed images (desired output). The network architecture consists of a feedforward deep neural network consisting of fully connected layers with hyperbolic tangent activations (which learns the transform), followed by convolutional layers with rectifier nonlinearity activations that form a convolutional autoencoder (which performs image domain refinement). Unfortunately, this results in a large number of parameters, which grows quadratically with the number of image pixels, which limited the use of AUTOMAP to small images (up to 128x128).

To reduce the parameter complexity of AUTOMAP, it is possible to decompose the two-dimensional inverse Fourier Transform into two one-dimensional iFTs; dAUTOMAP (decompose AUTOMAP) [42]. Here the model parameter complexity only increases linearly with the number of image pixels. A similar approach reduces the complexity using a multi-layer perceptron network to learn the one-dimensional iFT in a line-by-line approach, rather than the whole image [43, 44]. Alternatively it is possible to replace the fully connected layers of AUTOMAP, by a bidirectional RNN: ETER-net (End to End MR Image Reconstruction Using Recurrent Neural Network) [45]. ETER-net also decomposes the two-dimensional iFT, using two sequential recurrent neural networks. These methods showed a reduced number of training parameters, which allows its use in reconstruction of higher resolution images.

## 2.4 Cross-Domain Methods

Cross-domain methods are hybrid methods that operate in both the image domain and the frequency domain. They are based on the idea that CNN's operating on *k*-space and images exhibit different properties; therefore, a combination of them might outperform them separately. Typically, frequency domain subnetworks attempt to estimate the missing *k*-space samples, while image domain subnetworks attempt to remove residual artefacts.

Some cross-domain methods apply a single *k*-space completion step, followed by an image restoration step. This is the case of W-Net [46], which applies a frequency domain U-Net followed by an image domain U-Net. Another example is the multi-domain CNN (MD-CNN) [47], which uses a ResNet architecture for the *k*-space subnetwork and a U-Net for the image subnetwork in a dynamic imaging context.

Other hybrid methods use a cascading approach. In KIKI-Net [48], alternating *k*-space and image deep CNN's are applied, separated by the Fourier transform (the network architecture operates on *k*-space, image-space, *k*-space, and then image-space sequentially). Another proposal, the hybrid cascade [49], is based on a deep cascade of CNN's (DC-CNN) [47, 50]. However, unlike the original DC-CNN, it uses both *k*-space and image CNN's. The W-Net method was extended to WW-Net [51] by cascading more U-Net networks. This work also suggests that dual-domain networks may be most advantageous in multi-channel settings, where the *k*-space correlations between coils can be efficiently exploited by *k*-space domain networks.

A different approach is that of the dual-domain deep lattice network (DD-DLN) [52]. This method employs two DC-CNN's, one for each domain, which run in parallel rather than sequentially. In order to share information between both subnetworks, at the end of each block the outputs are concatenated (after transforming to the relevant domain) and fed into the next block in the cascade.

Another proposal is the Dual-Encoder-Unet [53], which unlike other methods is not based on single-domain subnetworks. Instead, a modified U-Net operates simultaneously on both domains, which is achieved by adding a second encoder path. One is fed the measured *k*-space, while the other is fed the zero-filled reconstructed images. The features from both paths are combined via concatenation and fed into a single decoder path, which produces the reconstructed image.

Finally, hybrid methods may operate on domains other than the image and the *k*-space domain. This is the case of IKWI-Net [54], which also includes a subnetwork in the wavelet domain (sequentially utilizing CNN's in the image domain, *k*-space, wavelet domain and image domain).

## 2.5 Unrolled Optimization

Unrolled optimization methods are inspired by iterative optimization algorithms used in compressed sensing MRI. The idea is to unroll the iterations of such an algorithm to an end-to-end neural network, mapping the measured *k*-space to the corresponding reconstructed image. Then image transforms, sparsity-promoting functions, regularization parameters and update rates can be treated as either explicitly or implicitly trainable and fitted to a training dataset using back-propagation. This has three advantages with respect to classic optimization. First, learned parameters may be better adapted to image characteristics than hand-engineered ones. Second, it avoids the need for manual tuning, which is not a trivial process. Finally, reconstruction is faster, because such learned iterative schemes are trained to produce results with fewer iterations.

Several optimization algorithms have so far been successfully unrolled into neural networks. These include gradient descent (GD) [55], proximal gradient descent (PGD) [50, 56, 57], the iterative shrinkage-thresholding algorithm (ISTA) [58], the alternating minimization algorithm (AMA) [59-61], the alternating direction method of multipliers (ADMM) [62, 63], and the primal dual hybrid gradient (PDHG) [64]. All unrolled methods solve some form of the following optimization problem:

$$\hat{x} = \operatorname*{argmin}_{x} f(Ax, y) + g(x) \qquad (4)$$

where $f(Ax, y)$ is a generic data consistency term, which ensures that the solution $x$ agrees with the observations $y$, and $g(x)$ is a generic regularization term which incorporates prior information. The definitions of $f$ and $g$, together with the optimization strategy, determine the fundamental structure of the resulting neural network. Several approaches are outlined hereafter. A summary of the techniques described is presented in Table 1, which the reader is encouraged to use for reference.

| Ref. | Name | Algorithm | $f$ | $g$ | Learned parameters |
|---|---|---|---|---|---|
| [63, 65] | ADMM-Net | ADMM | $\frac{1}{2}\|Ax-y\|_2^2$ | $\sum_{l=1}^{L}\lambda_l \mathcal{R}(D_l x)$ | $D_l$ (Conv), $\mathcal{R}$ (implicit, proximal operator, piecewise linear function) |
| [55] | VarNet | GD | | | $D_l$ (Conv), $g$ (implicit, first order derivative, radial basis functions) |
| [58] | ISTA-Net | PGD (ISTA) | | $\lambda\|D(x)\|_1$ | $D$ (Conv-ReLU-Conv) |
| [57] | R-GANCS | PGD | | $\mathcal{R}(x)$ | $\mathcal{R}$ (implicit, proximal operator, GAN). |
| [56] | HC-PGD | PGD | | $\mathcal{R}(x)$ | $\mathcal{R}$ (implicit, proximal operator, CNN). |
| [50] | DC-CNN | PGD | | $\lambda\|x-\mathcal{C}(x)\|_2^2$ | $\mathcal{C}$ (CNN) |
| [59] | MoDL | AMA | | | $\mathcal{C}$ (CNN) |
| [66] | MoDL-SToRM | AMA | | $\lambda_1\|x-\mathcal{C}(x)\|_2^2 + \lambda_2\,\mathrm{tr}(x^T L x)$ | $\mathcal{C}$ (CNN) |
| [60] | VS-Net | AMA | | $\mathcal{R}(x)$ | $\mathcal{R}$ (implicit, proximal operator, CNN). |
| [61] | CRNN-MRI | AMA | | $\mathcal{R}(x)$ | $\mathcal{R}$ (implicit, proximal operator, CRNN). |
| [64] | TVINet | PDHG | | $\mathcal{R}(Dx)$ | $D$ (conv), $\mathcal{R}$ (CNN) |
| [67] | PDHG-CSNet | PDHG | | $\mathcal{R}(x)$ | $\mathcal{R}$ (implicit, proximal operator, CNN). |
| [67] | CP-Net, PD-Net | PDHG | $\mathcal{F}(Ax,y)$ | $\mathcal{R}(x)$ | $\mathcal{F}$, $\mathcal{R}$ (implicit, proximal operators, CNN's) |

*Table 1: Summary of unrolled optimization methods. A selection of unrolled optimization methods and their fundamental characteristics: optimization algorithm, data consistency term, regularization term and learned parameters. Regularization parameters λ, as well penalty parameters, update rates, step sizes, etc., are learned too, but omitted from the learned parameters column for conciseness. ADMM: alternating direction method of multipliers; GD: gradient descent; PGD: proximal gradient descent; ISTA: iterative shrinkage-thresholding algorithm; AMA: alternating minimization algorithm; PDHG: primal dual hybrid gradient method; Conv: convolutional layer; ReLU: rectified linear unit; GAN: generative adversarial network; CNN: convolutional neural network; CRNN: convolutional recurrent neural network. In the regularization term for MoDL-SToRM, tr denotes the trace operator and L denotes the graph Laplacian operator.*

Like their compressed sensing counterparts, most unrolled reconstruction methods define data consistency in the least-squares sense, assuming that measurement noise is normally distributed:

$$f(Ax, y) = \frac{1}{2}\|Ax - y\|_2^2 \tag{5}$$

There exists more variability in the use of regularization functions. Some of the earliest approaches consider a regularization term of the form $g(x) = \mathcal{R}(Dx)$, which contains an explicit sparsifying transform $D$ and a sparsity-promoting function $\mathcal{R}$. This formulation is similar to compressed sensing, where $D$ might be the wavelet transform or the finite difference operator, and $\mathcal{R}$ would typically be the $\ell_1$-norm. In unrolled optimization, these terms can be learned rather than manually designed. ADMM-Net [63, 65], VarNet (Variational Network) [55] and TVINet (Total Variation Inspired Network) [64], which unroll ADMM, GD and PDHG, respectively, use this formulation. All three explicitly learn linear sparsifying transforms $D$, parameterized by convolutional layers, and non-linear sparsity-promoting functions $\mathcal{R}$. The latter are not learned directly, but rather implicitly through their proximal operators. Different parameterizations are used: ADMM-Net uses piecewise linear functions, VarNet uses radial basis functions, and TVINet uses a CNN. Another method, ISTA-Net (Iterative Shrinkage-Thresholding Algorithm) [58], uses the regularizer $\lambda\|D(x)\|_1$, where $D$ is a non-linear sparsifying transform (two convolutional layers separated by a ReLU activation). In this case, the sparsity-promoting function is not learned, but fixed to be the $\ell_1$-norm.

Another class of methods, inspired by image restoration approaches (see section 2.1), use the regularizer $g(x) = \|x - \mathcal{C}(x)\|_2^2$, designed to formulate an explicit image denoising problem. Here $\mathcal{C}$ is an operator that removes noise and aliasing artefact from an image. As a result, the overall term $g(x)$ is a noise estimator. Naturally, the operator $\mathcal{C}$ is complex and unknown, but can be learned by a CNN. Methods using this approach include DC-CNN (a Deep Cascade of CNN's) [50] and MoDL (Model-Based Deep Learning) [59].

Some work has combined several regularizers in the same network. It is the case of MoDL-SToRM (MoDL with SmooThness regularization on manifolds) [66], which combines a learned noise estimator with a fixed SToRM regularizer. The first operates as has just been discussed, while the latter ensures that the reconstructed dynamic sequence lies in a smooth low-dimensional manifold.

Finally, some methods do not constrain the formulation of the regularizer. Instead, they consider a generic term $\mathcal{R}(x)$, and use a CNN to estimate its proximal mapping directly. This is the case of R-GANCS [57], CRNN-MRI (convolutional recurrent neural network) [61], VS-Net (Variable Splitting Network) [60], HC-PGD (history cognizant PGD) [56] and PDHG-CSNet (primal dual hybrid gradient, compressive sensing) [67].

Although most methods use the $\ell_2$-norm as the data consistency function, other approaches have been proposed. In CP-Net (Chambolle-Pock Net) [67] the data consistency term is relaxed to a generic form $f(x) = \mathcal{F}(Ax, y)$, where $\mathcal{F}$ is learned. Similar relaxations have been proposed for ADMM and ISTA-based unrolled networks [67, 68]. This may increase the generality of the models, at the cost of looser data consistency guarantees.

Some methods [67-69], such as PD-Net (Primal Dual Net), also suggest relaxing the update rules, which are otherwise determined by the optimization algorithm, to further increase the generality of the model. A related approach is that taken in recurrent inference machines (RIMs) [70-72]. These parameterize the optimization process as a recurrent neural network, where each "time step" is an iteration of the optimizer. RIMs learn the optimizer itself along with the prior; therefore, unlike other approaches in this section, they are not based in any particular optimization algorithm. The underlying idea is that a specialized data-driven optimizer might outperform hand-designed ones. Unrolled optimization methods can incorporate parallel imaging by making coil sensitivity operators a part of the image formation model $A$. Some of the approaches outlined here have done so [56, 59, 60, 63], while others have worked in a single-coil context [50, 58, 61, 65, 67]

The different formulations and optimization algorithms lead to significant variability in the resulting network architectures, which cannot be covered in detail in this review. However, some structures are found often. For example, proximal gradient methods [50, 56, 57] and alternating minimization methods with quadratic penalty splitting [59, 61] map naturally to alternating blocks in the resulting neural network: a model-driven data consistency block and a learned prior block (see Figure 3). Augmented Lagrangian methods, such as ADMM-Net [62], exhibit in addition an update block for the Lagrange multiplier.

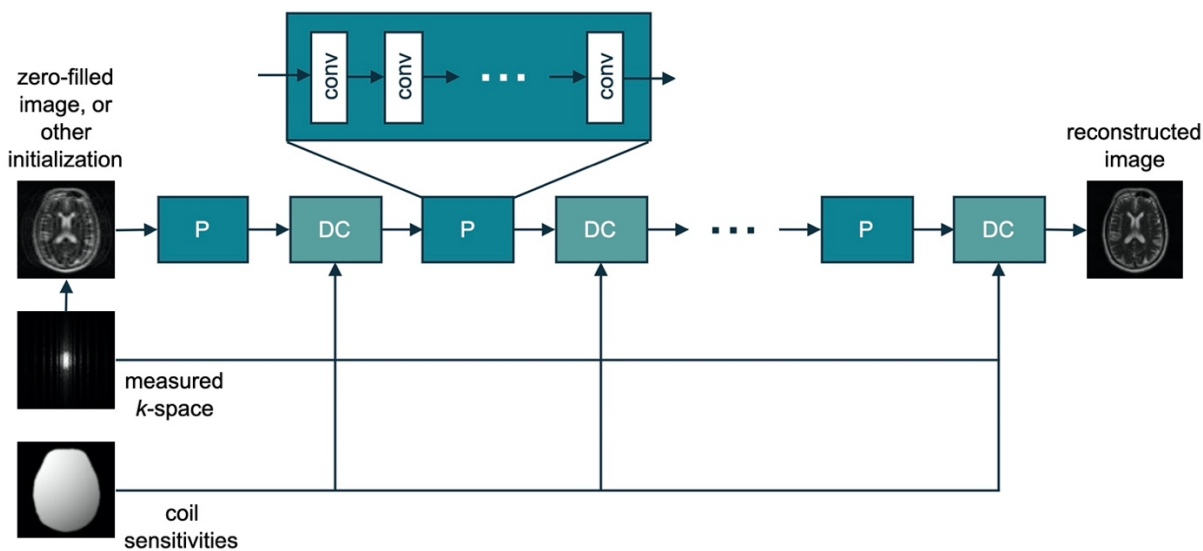

*Figure 3: Example of an unrolled optimization deep neural network, with an alternating structure containing data consistency (DC) blocks and prior (P) blocks. DC blocks implement a gradient descent or proximal mapping step to minimize the data consistency term. They use the system matrix A and the original k-space measurements, and may use coil sensitivities in multi-channel settings. P blocks implement the proximal mapping of the regularization term and are learned by a CNN. The exact CNN architectures vary between methods. Note that this is not an accurate representation of all unrolled networks, but it shows commonly found features and is the basic backbone of several of the methods presented.*

All unrolled methods are ultimately deep neural networks, with the important difference, with respect to other deep learning approaches, that their architecture is informed by a physics-driven model. There are also important differences with respect to traditional optimization methods, besides the obvious data-driven design. For example, they are often truncated to a fixed number of steps (iterations), and trained in an end-to-end fashion, with backpropagation across steps. They may share weights across steps [59, 61], or each step may have its own weights [50, 55], thus imparting different behavior to different steps. There may be additional components which do not have an immediate optimization equivalent, often borrowed from the rich body of deep learning literature. For example, in R-GANCS [57], GAN's are used to enhance the perceptual quality of the reconstructed images. In CRNN-MRI [61], recurrent units are used to exploit redundancies across iterations as well as along the dynamic dimension. In HC-PGD [56], dense connections are added across steps in order to accelerate convergence and improve overall performance.

As a result of all this variability, unrolled methods may rely on training data to different extents. Model-driven approaches might use a more constrained formulation and shallower priors, and have a smaller number of parameters [55, 58]. Such methods may be easier to interpret and validate, and may require less training data. As constraints are relaxed and deeper priors are used [59, 67], methods become more data-driven and may have more parameters. Such methods have a looser connection to the physics-driven model, but given enough training data they might outperform more model-driven approaches on a particular task, due to increased representational power.

## 3 Unsupervised Machine Learning

In unsupervised learning the ML algorithm looks to find patterns in data without the need for any *ground-truth* data or user guidance. This is particularly challenging in the field of MRI reconstruction. It has been shown that state-of-the-art unsupervised learning techniques are currently unable to achieve as good image quality as supervised learning techniques [73, 74]. However, in applications where *ground-truth* fully sampled datasets are unavailable and difficult or impossible to acquire (e.g. 4D flow), unsupervised learning techniques provide a promising alternative.

Unsupervised learning has been used to train image restoration methods to remove noise from MRI images (see section 2.1) using only noisy training data; examples include Noise2Noise [75] and regularization by artifact-removal (RARE) [76]. Additionally, unsupervised learning is used in DeepResolve [77], in which a 3D cascade of convolutional filters is trained to perform super-resolution (see section 2.1).

Other unsupervised approaches which have shown promise, are algorithms which exploit image sparsity, similarly to compressive sensing. These simultaneously reconstruct the image and learn dictionaries or sparsifying transforms for image patches (also called *blind compressed sensing*) [78, 79]. A further extension to this is Deep Basis Pursuit (DBP), which uses known noise statistics for each data set. This unrolled optimization alternates between auto-encoder CNN layers and data consistency constraint of basis pursuit de-noising [74]. Actual data consistency has also been used by cross-validation [80].

Generative adversarial networks have been used to enforce data consistency in unsupervised learning. Here, a conditional GAN is used to directly learn the mapping from *k*-space to image domain [73], where the generator network outputs an image (from undersampled *k*-space data), and the discriminator network tries to differentiate between the original *k*-space and a randomly undersampled *k*-space created from the generated image. GAN's have also been used to learn the probability distribution of uncorrupted MRI data in an unsupervised manor, and provide implicit priors for iterative reconstruction approaches [81]. Superior image quality may be achieved by also allowing the generating network to learn its range space with respect to the measured data [82].

## 4   Clinical Implications

Despite the number of publications showing technical advances in machine learning for MRI reconstruction, many publications do not demonstrate clinical utility. Instead performance of the resulting network is often evaluated using quantitative metrics generated from synthetic data, including MSE, MAE, Root Mean-Squared Error (RMSE), Peak signal-to-noise ratio (pSNR) and Structural Similarity Index (SSIM). However, these metrics do not agree well with expert radiologists in ascertaining image quality, and ultimately diagnostic confidence [5]. In addition, real data may not perform as well as synthetic data, therefore demonstration in prospective data sets is essential. In order to move towards clinical translation, it is necessary to evaluate qualitative image quality, diagnostic scoring and measurement of quantitative clinical metrics (against reference standard imaging techniques) from prospectively acquired data reconstructed using ML.

There have been a small number of clinical validation studies of ML reconstructions, in particular within cardiovascular MRI. In one study, real-time acquisition of 2D cine data was achieved using a radially 13x undersampled acquisition, with a ML de-aliasing reconstruction (see section 2.1) [14]. After training of the network, prospective data was acquired in 10 patients with Congenital Heart Disease (CHD) and reconstructed using the ML network. Qualitative image scoring (myocardial delineation, motion fidelity, and artefact) and clinical measures of left and right ventricular volumes were compared to those from clinical gold-standard images. No statistically significantly differences were found in qualitative image quality or left ventricular volumes (EDV, end diastolic volume; ESV, end systolic volume; and EF, ejection fraction), with a small underestimation of right ventricular end systolic volume (bias -1.1 mL). This study demonstrated a reduction in total scan time from ~279 s for gold-standard acquisition to just ~18 s, where the ML reconstruction was >5x faster than a CS reconstruction of the same data.

Another study quantified left ventricular volumes in 20 healthy subjects and 15 patients with suspected cardiovascular disease, from a 3D CINE sequence with an unrolled ML network: CINENet [83] (which resembles a proximal gradient algorithm with sparsity-learning and data consistency steps). This also found good agreement in LV function ESV, EDV and EF compared to clinical gold-standard images, enabling 3D CINE data to be acquired in less than 10 s scan with ~5 s reconstruction time.

In another clinical validation paper, vessel diameters, diagnostic accuracy and diagnostic confidence were assessed from 3D whole-heart images with a single volume super-resolution ML reconstruction (see section 2.1) [27]. Prospective data was acquired in 40 patients with CHD, and compared to results from clinical gold-standard images. Qualitative image scoring showed super-resolved images were similar to high-resolution data (in terms of edge sharpness, residual artefacts and image distortion), with significantly better quantitative edge sharpness and signal-to-noise ratio. Vessel diameters measurements showed no significant differences and no bias was found in the super-resolution measurements in any of the great vessels. However, a small but significant for the underestimation was found in coronary artery diameter measurements from super-resolution data. Diagnostic scoring showed that although super-resolution did not improve accuracy of diagnosis compared to low-resolution data, it did improve diagnostic confidence. This study demonstrated a ~3x speed-up in acquisition compared to high-resolution data (173 s vs 488 s), where super-resolution reconstruction took < 1 s per volume.

Vessel diameters have also been quantified from four-dimensional non-contrast MRI angiography data, with a ML de-aliasing reconstruction (see section 2.1) in 14 patients with thoracic aortic disease [84]. Unfortunately, comparisons were made against a CS reconstruction, rather than a gold-standard technique, but showed clinically acceptable visual scores, with no significant difference in terms of mean vessel diameters for six out of seven standardized locations in the thoracic aorta. In another study, coronary artery length has been measured from ML reconstructed 3D angiographic data (using a multi-scale variational neural network, see section 2.5) [85]. They showed negligible differences in terms of quantitative vessel sharpness and coronary length, compared to a fully-sampled scan in 8 healthy subjects.

Myocardial scar quantification has been performed for 3D late gadolinium enhancement (LGE) MRI data reconstructed with a ML de-aliasing reconstruction (see section 2.1) [86]. Unfortunately, the study compared the ML reconstructed data against a CS reconstruction, rather than a gold-standard technique, however an excellent correlation in scar extent was observed (with a per-patient scar percentage error was 0.17 ± 1.49%).

Flow quantification has been calculated from 2D phase contrast data in 14 subjects, with a *k*-space interpolation ML reconstruction (see section 2.2) [15]. Unfortunately, the data was retrospectively undersampled, however the flow waveforms and flow volumes were seen to agree well with fully-sampled data, although the acceleration rates were low (x2, x3 and x5). Another study extended this to 4D flow using a deep variational neural network to perform an unrolled reconstruction (see section 2.5) [87]. The resultant network was tested on prospectively on 7 healthy subjects, and compared to a gold-standard technique, with good agreement in terms of peak-velocities and peak-flow estimates.

## 5  Current Limitations

Raw MRI data is complex-valued, however many machine learning frameworks do not use complex convolutions or complex activation functions. Some studies just use magnitude data (particularly in image restoration methods), whereas others train separate networks for the magnitude and phase data [88], or may separate the real and imaginary parts into two separate channels [34, 50, 55]. These approaches do not necessarily maintain the phase information of the data. Development of complex-valued networks remains an area of active research [89-91]. However, PyTorch has recently (year: 2020) introduced full complex value support, which means that more studies may use complex-valued data in the future.

Many studies only consider single channel data, whereas raw data is normally acquired from multiple coils. Some studies handle multi-coil data without additional coil-sensitivity information or ACS lines [92], whereas others learn the coil weighting from ACS lines in training [36], and some feed pre-calculated coil sensitivity maps into the network [55].

There is a question about how specific a network needs to be. Even where imaging is fixed to a specific anatomy, the image quality can be variable. This may be due to different hardware (including field strength and coils), the use of different protocols (including different imaging contrasts, acquisition trajectories, flip angles, bandwidth and pre-pulses), patient-specific variation (including different field-of-view, matrix size, phase-encoding direction), and artefacts (e.g. from patient motion). There may also be great variability in the prescribed scan planes, as well as in the underlying anatomy across different diseases. As most articles report their results on private data sets, it is difficult to compare the methods and assess their robustness and generalizability.

Currently there are only a relatively small number of publicly available data sets, and these are often very specific. These include (but are not limited to) raw *k*-space data sets; mridata.org, NYU fastMRI [93] and Calgary-Campinas-359 [94], as well as DICOM imaging data sets; UK Biobank [95], Hunan Connectome Project [96], The Montreal Neurological Institute's Brain Images of Tumors for Evaluation (NTI BITE) [97] and OASIS-3 [98]. The availability of these datasets enables development of novel DL image reconstruction frameworks, as well as making it possible to benchmark and compare networks in the same setting [99].

One of the main limitations to successful use of machine learning reconstructions in MRI is the lack of integration into the clinical environment. This means that currently reconstructions are performed off-line and are not available immediately to the clinician. Manufacturers have been working to integrate machine learning frameworks into standard clinical pipelines. In addition open source frameworks which may be integrated into the scanner, such as Gadgetron [100], may also enable translation of these techniques into the clinical environment. This would also enable large multi-site validation studies to be performed, which is essential in building confidence in these techniques.

## 6 Conclusion

Deep learning approaches have been shown to provide a huge potential for the future of magnetic resonance image reconstruction. There has been an explosion of research in this field over the last five years, across many different approaches. More robust testing and large-scale demonstration on prospectively acquired clinical data is required to build confidence in these techniques.

**Funding sources**

JAS and JMT are funded under the UKRI Future Leaders Fellowship (MR/S032290/1). JMT is also part funded by Heart Research UK (RG2661/17/20). This work was supported in part by British Heart Foundation grant: NH/18/1/33511. AH is funded by Academy of Finland Projects 312123 (Finnish Centre of Excellence in Inverse Modelling and Imaging, 2018–2025).

**Highlights**

- Machine learning reconstruction of MRI data is becoming increasingly popular in research
- Many methods exist to perform machine learning reconstruction of MRI data
- The limited availability of publicly available training data sets, restricts current development and comparison of existing methods

- There is currently very limited clinical validation of MRI images reconstructed using machine learning